4DTIP of MEMS & MOEMS 9-11 April 2008

# Processing and Characterization of Precision Microparts from Nickel-based Materials

D. M. Allen[1], H.J. Almond [1], K. Bedner [1,2], M. Cabeza [1], B. Courtot [1], A. Duval [1], S.A. Impey [1] and M. Saumer [2]

[1] Department of Advanced Materials, Cranfield University, Bedford MK43 0AL, UK
[2] Kaiserslautern University of Applied Sciences, D-66482 Zweibrücken, Germany

*Abstract-* The objective of this research was to study the influence of electroplating parameters on electrodeposit characteristics for the production of nickel (Ni) and nickel-iron (Ni-Fe) microparts by photoelectroforming. The research focused on the most relevant parameter for industry, which is the current density, because it determines the process time and the consumed energy. The results of the Ni and Ni-Fe characterisations can be divided into two aspects closely linked with each other; the morphology and the hardness.

I. INTRODUCTION

The process of photoelectroforming (PEF), also known as photoforming [1], now plays an important role in the fabrication of metal microparts. The process employs photoresist technology to coat and pattern a conductive mandrel. The apertures in the resist stencil are then filled with an electrodeposited material (typically nickel-, silver-, gold- or copper-based). The bond between the metal and the mandrel is designed to be weak so that the electrodeposited material can be easily removed from the pre-treated mandrel surface by flexure of the mandrel. The resultant electroform then comprises a component in its own right, rather than existing as an adherent surface film as seen in electroplated microparts.

The electroplating process results in the growth of a nanocrystalline material. One of the challenges of this project was to determine the morphology of the deposited Ni (as a baseline) and Ni-Fe samples as previous literature values do not appear to be consistent [2-6]. The value of a Ni-Fe deposit is that it is commonly used in microparts that are subjected to potential wear from frictional forces, such as those encountered by trains of micro-gearwheels. Therefore, the electroformed gearwheels need to be hard and durable, so that wear and fracturing is reduced to increase the operational lifetime of the gear train mechanism.

II CURRENT REGIMES

Three types of current regime can be used in PEF; direct (DC), pulsed and pulse reverse (PR) currents. The first regime is in very common use, with the other, more expensive, regimes being developed as the control technology and product knowledge base develop with time. In these experiments, only two current regimes were investigated; DC and PR. Ni samples of thicknesses between 100 and 200 µm were electroplated in a nickel sulphamate bath using a sulphur-activated nickel anode that prevents passivation. The bath composition comprised nickel sulphamate (400 g/l), boric acid (35 g/l), nickel chloride (4.3 g/l) and distilled water. The pH was held at 4.6 (an appropriate value for this process) and electroplating was carried out at a temperature of 50°C and with both constant and variable current densities in the range of 0.5 to 1 A/dm². The cathode comprised a 5 x 5 cm brass sheet.

The Ni-Fe micro components were electroformed in Zweibrücken with DC and PR (f=20 Hz) utilising current densities of 2 A/dm$^2$ (for 90 min) followed by 3 A/dm$^2$ (for 100 min) and, lastly, an additional 5 A/dm² (for 90 min) The alloy was electrodeposited over a silicon wafer coated with a 100nm thick layer of sputtered titanium and a 120 µm thick SU-8 photoresist pattern (Figure 1). The total thickness of the micro components was about 140 µm.

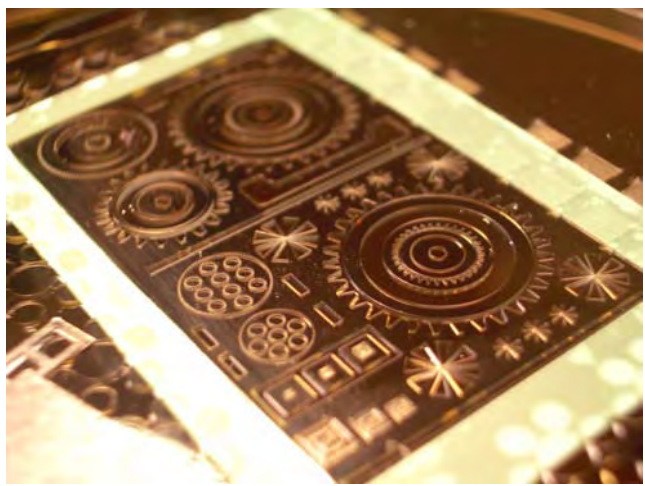

Fig.1. SU-8 photoresist mould on the conductive silicon wafer

III DEPOSIT MORPHOLOGY

Morphology was determined by two methods. The first method involved sectioning the parts by FIB and then examining the sections by scanning electron microscopy (SEM). The second method involved examining by optical microscopy after embedding the samples in conductive phenolic resin. Following sectioning, grinding and polishing, the deposits were etched in nitric acid / glycerol / hydrochloric acid (50:25:15 v/v) to enhance the visual contrast and grain definition.

The morphologies of the Ni-Fe DC samples differ from those of the Ni-Fe PR samples. In the latter, the morphology becomes coarser with an increasing iron content in the range of 17-29 at%, whereas the morphology of the DC sample is fine for iron contents between 28 and 32 at%. In the Ni DC sample the structure is columnar initially but becomes finer as the thickness develops from the brass mandrel through to the top of the deposit, most probably due to a mandrel effect (possibly catalysed by the grain size of the brass sheet on its surface).

In the case of an incremental step-change in current density during the electroplating process, the shape of the morphology changes from coarse columnar to fine equiaxed. This observation could not be repeated for the Ni-Fe DC sample because the grain size is so fine that the morphology looks completely uniform on examination by both optical microscopy and SEM.

During electroplating, the current density can be changed gradually or rapidly and this variable also has an influence on the resultant deposit. A rapid change results in a very sharp interface between two different morphologies as shown in Figure 2, whereas, in the case of a gradual change, only a transition zone is observed. Changes in morphology on the equivalent PR deposits are shown in Figures 3-5.

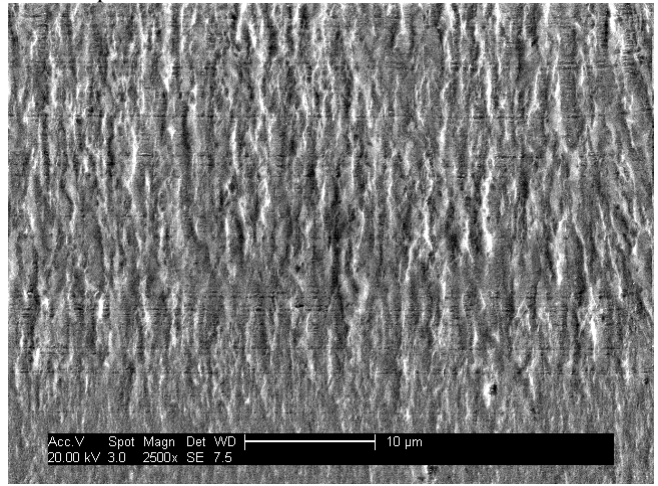

Fig. 3. Ni-Fe microstructure of the top surface PR deposited at <<5 A/dm$^2$ where the part has started to overplate outside the mould boundaries leading to a coarser morphology

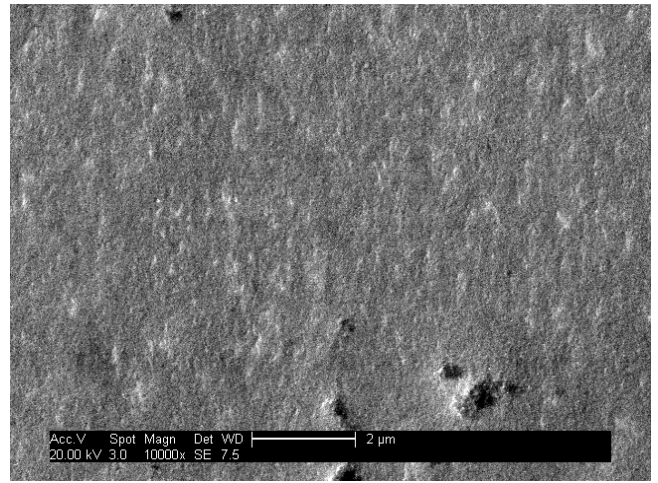

Fig. 4. Microstructure of Ni-Fe PR deposited at 3 A/dm$^2$ (Note the change in the scale bar dimension c.f Fig.5.

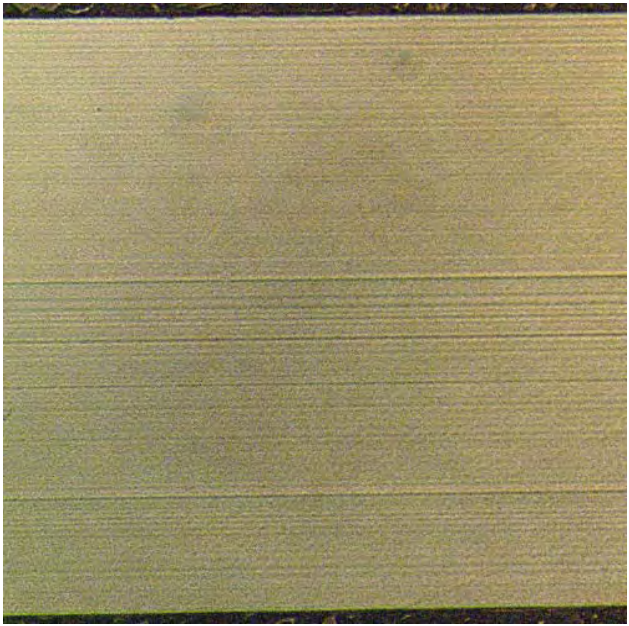

Fig.2. Cross-section of a Ni-Fe 170µm thick DC deposit (with the mandrel at the bottom) showing two prominent demarcation lines where the current density was abruptly changed from 2 to 3 A/dm$^2$ and then from 3 to 5 A/dm$^2$.

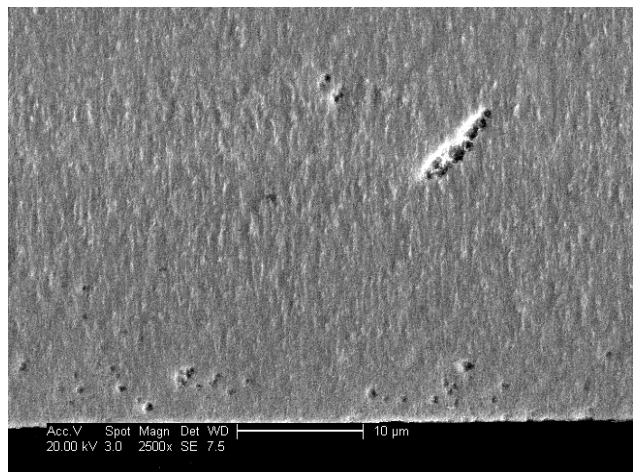

Fig.5. Microstructure of Ni-Fe PR deposited at 2 A/dm$^2$ on the surface of the mandrel

## IV GRAIN SIZE

The size of the grains in electrodeposited Ni-Fe films is difficult to evaluate, especially as we now know that the grain shape changes significantly with respect to current density. The grain size has been evaluated by several methods. FIB analysis gave very different answers compared to the literature and we believe that FIB only shows grain aggregates. On the other hand, examination by X-ray diffraction (XRD) and feeding of the results into the Scherrer equation (1) gave very plausible results where c (constant) = 0.9, λ = wavelength of the x-ray radiation, θ = glancing angle and FWHM = full width at half maximum of the peak.

$$D = c \cdot \frac{\lambda}{FWHM * \cos(\theta)} \quad (1)$$

Prior to the measurements, the mandrel side was polished to remove the material that might be influenced by the mandrel. The polished mandrel sides of both samples were analysed with a Siemens D 5005 x-ray diffractometer having a copper x-ray tube with a wavelength of 1.54 Å. After the measurements, the mandrel sides of the samples were subjected to energy dispersive X-ray analysis (EDX) to determine the iron content of the region examined previously with the x-ray diffractometer. The results of the XRD measurements are shown in Figure 6 and exhibit two distinct peaks, the (111)-peak and the (200)-peak. These peaks are characteristic for nickel. This indicates that the face-centred cubic (fcc) structure of nickel is preserved with the iron atoms implanted within the nickel lattice. It has been shown previously that the iron content has an influence on the crystal structure as it changes from the fcc-structure of nickel to the body-centred cubic (bcc) structure of iron when the nickel-iron alloy has an iron content larger than 53 at% [7]. For materials arranged in a cubic system the lattice parameter $a_0$ can be calculated from equation (2):

$$a_0 = d_{hkl} \cdot \sqrt{h^2 + k^2 + l^2} \quad (2)$$

where d = d-spacing and h, k and l are the Miller's indices. The lattice parameters and grain sizes for DC and PR Ni-Fe are shown in Table 1.

## V HARDNESS OF THE DEPOSITS

The application of photoelectroforming to the fabrication of micro gear wheels requires a hard material for longevity of gear train components and therefore the effect of the current density on the hardness of Ni and Ni-Fe was investigated and measured with a pendulum nanoindenter (Micro Materials Nanotest 600).

For the DC samples, the hardness is constant. All the Ni DC samples investigated had a hardness value between 3.5 and 4.5 GPa. No relationship of hardness to current density was apparent in a range of current densities from 0.5 to 1.0 A/dm².

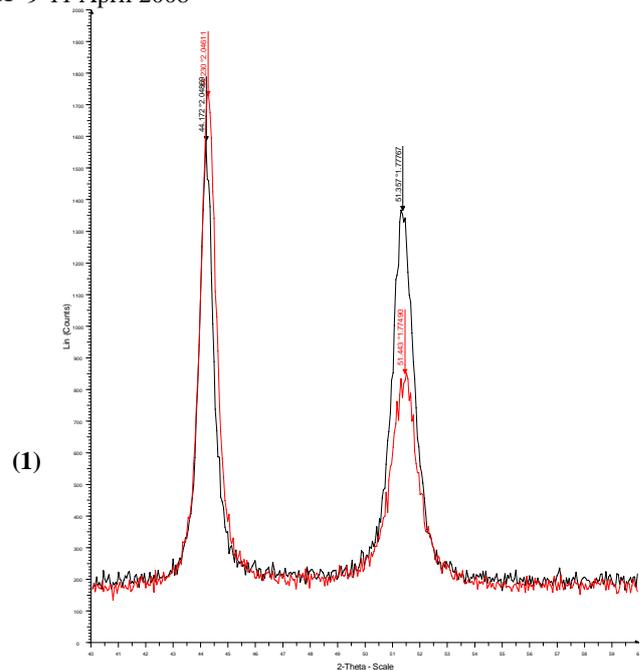

Fig.6. Pattern of the XRD measurement of the dc sample (red) and the 20 Hz sample (black). The plot shows the (111) peaks at 44.230° and 44.172° for the dc sample and the 20 Hz sample, respectively. The (200) peaks are at 51.443° and 51.357° for the dc sample and the 20 Hz sample, respectively. Both peaks of the dc sample are shifted to the left, indicating a slight stress in the lattice.

For the Ni-Fe DC samples, the hardness has a constant value of about 7 GPa within an iron content of 28 to 32 at%. A comparison of the results for the DC samples shows that the Ni-Fe alloy is harder than the pure Ni: the iron-alloying increasing the hardness of the deposit.

The pulse reverse (PR) samples are more complex because the hardness is influenced both by the morphology and the iron content as shown in Fig.7. It is difficult to separate the influence of the morphology from that of the iron content as both affect each other. In the range of 17 to 23 at% of iron, the morphology is fine-grained and from 23 to 29 at% iron it is coarse-grained. The zone of fine morphology has a hardness of 5.0 to 7.0 GPa whereas the coarse material is softer with a hardness of 3.5 to 5.0 GPa.

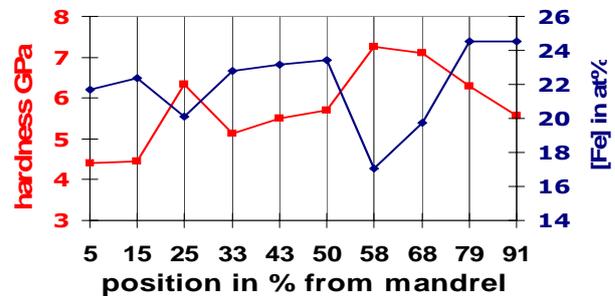

Fig. 7. An analysis of the hardness of deposit with respect to at% of iron in a typical PR Ni-Fe deposit

Table 1. Lattice parameters and grain sizes for the DC and PR Ni-Fe samples

|  |  | lattice parameter a in Å | grain size D in nm of (111) peak | intensity ratio (200)/(111) | iron content in wt% |
|---|---|---|---|---|---|
| dc sample | (111) reflection | 3,5440 | 13 | 0.49 | 17.27 |
| dc sample | (200) reflection | 3,5498 |  |  |  |
| 20 Hz sample | (111) reflection | 3,5484 | 15 | 0.86 | 17.76 |
| 20 Hz sample | (200) reflection | 3,5553 |  |  |  |

In summary, the hardness relates to the inverse of the iron content in the range of alloys examined. The influence of current density is complex because it controls not only the iron content but also the morphology and therefore the hardness.

## VI CONCLUSIONS

The results obtained make it possible to give some guidelines to industry in order to improve the performance and longevity of photoelectroformed Ni-Fe micro gearwheels .

In terms of maximum hardness, the products could be

- Ni-Fe plated with DC, giving a constant hardness of 7GPa with an approximately constant iron concentration of 30 at%, but internal stresses appear with DC samples that frequently peel off from the mandrel and curl. The resultant product therefore cannot be used in a practical application!
- Ni-Fe plated with PR current, which has a hardness of 5-7 GPa and an iron content in the range 17-23 at%, to produce less stressed samples but requiring a slightly higher deposition time (approximately 5% higher at a rate of 0.2 µm/Adm$^{-2}$min) and associated increased costs.

By using high current density it was determined that

- A fine morphology which gives a higher hardness also shortens the process time resulting in a cheaper process (less energy applied, higher production rate) but the influence of current density on the alloying element concentration needs to be borne in mind.
- Grain size was approximately 13nm for DC deposits and 15nm for PR deposits

To avoid non-uniformity in morphology, any change in current density whilst electroplating should be gradual, to avoid sharp interfaces between different morphologies where cracks and fractures might concentrate when the part is stressed.

Future work will be concentrated on determining good DC / pulse / pulse reverse parameters using high current density and yet still maintaining homogeneous material properties.

## ACKNOWLEDGMENT

D.M.A., H.J.A. and S.A.I. wish to thank K.B. for her invaluable technical liaison with S.M., Dr Tony Hart (Hart Coating Technology Ltd, UK) for technical discussions on nickel electroforming and EPSRC for Grant EP/C534212/1 to study "3D-Mintegration" that involved development of a demonstrator nickel electroplated metrology microprobe.